\documentclass[letterpaper,arxiv]{dae-handout}
\graphicspath{{./}{./FIGURES/}{./figures/}{../Figures/}{../../Figures/}{./images/}{./graphics/}}

\newcommand{\trversion}{v5.0}
\newcommand{\trfilename}{Lamports-Arrow-of-Time.tex}
\newcommand{\trdate}{02026-FEB-25}
\newcommand{\trshorttitle}{Lamport's Arrow of Time}
\newcommand{\trauthor}{Paul Borrill}
\newcommand{\traffiliation}{D\AE D\AE LUS}

\usepackage[T1]{fontenc}
\usepackage[utf8]{inputenc}
\usepackage[bitstream-charter]{mathdesign}
\DeclareSymbolFont{operators}   {OT1}{cmr} {m}{n}
\DeclareSymbolFont{letters}     {OML}{cmm} {m}{it}
\DeclareSymbolFont{symbols}     {OMS}{cmsy}{m}{n}
\DeclareSymbolFont{largesymbols}{OMX}{cmex}{m}{n}
\SetSymbolFont{operators}{bold} {OT1}{cmr} {bx}{n}
\SetSymbolFont{letters}  {bold} {OML}{cmm} {b}{it}
\SetSymbolFont{symbols}  {bold} {OMS}{cmsy}{b}{n}
\SetMathAlphabet{\mathit} {normal}{OT1}{cmr}{m}{it}
\SetMathAlphabet{\mathbf} {normal}{OT1}{cmr}{bx}{n}
\SetMathAlphabet{\mathsf} {normal}{OT1}{cmss}{m}{n}
\SetMathAlphabet{\mathtt} {normal}{OT1}{cmtt}{m}{n}
\usepackage{amsmath}
\usepackage{amsthm}
\usepackage{booktabs}
\usepackage{enumitem}
\usepackage{fancyhdr}
\usepackage{titletoc}
\usepackage{etoc}
\usepackage{lastpage}
\usepackage{xcolor}
\usepackage{tikz}
\usepackage{eso-pic}
\usepackage{url}
\usepackage{morefloats}   
\usepackage{placeins}     
\usetikzlibrary{positioning,shapes.geometric}

\setcitestyle{numbers,square}

\newtheorem{theorem}{Theorem}[section]

\newtheorem{definition}[theorem]{Definition}

\newcommand{\fito}{\textsc{fito}}
\newcommand{\tlaplus}{\textsc{TLA}$^{+}$}
\newcommand{\paxos}{\textsc{Paxos}}

\tikzset{
  badge/.style={
    circle,
    draw=#1,
    fill=#1!10,
    line width=0.5pt,
    minimum size=0.45in,
    font=\tiny\sffamily,
    align=center,
    text=#1!80!black
  }
}

\newcommand{\placebadges}{%
  \ifarxiv\else
  \AddToShipoutPictureBG*{%
    \AtPageUpperLeft{%
      \raisebox{-0.5in}{\hspace{\dimexpr\paperwidth-0.7in\relax}%
        \begin{tikzpicture}[overlay]
          \node[inner sep=0pt] (logo)
            {\includegraphics[height=0.855in]{DAE-Logo.png}};
          \node[badge=green!60!black, left=0.08in of logo]  (b1) {Artifacts\\Available};
          \node[badge=purple!70!black, left=0.08in of b1]   (b2) {Expert\\Verified};
          \node[badge=green!50!black,  left=0.08in of b2]   (b3) {AI\\Assisted};
          \node[badge=blue!70!black,   left=0.08in of b3]   (b4) {Human\\Conceived};
        \end{tikzpicture}%
      }%
    }%
  }%
  \fi
}

\newcommand{\maketrcover}{%
  \ifarxiv\else
  \thispagestyle{empty}
  \begin{fullwidth}
  \vspace*{2in}
  \begin{center}
    {\Large\sffamily\bfseries D\AE D\AE LUS Technical Report}\\[1.5em]
    {\LARGE\sffamily\bfseries Lamport's Arrow of Time:\\[0.3em]
     The Category Mistake in Logical Clocks}\\[2em]
    {\large \trauthor\,,\;\traffiliation}\\[1em]
    {\normalsize \trversion\quad---\quad\trdate}
  \end{center}

  \vspace{2em}
  \noindent\rule{\linewidth}{0.4pt}

  \vspace{1em}
  \footnotesize
  \begin{description}[leftmargin=1.2in, style=sameline, font=\normalfont\scshape]
    \item[Status:]       v5.0 --- arXiv submission baseline
    \item[Filename:]     \texttt{\trfilename}
    \item[Keywords:]     Lamport, happens-before, logical clocks, arrow of time,
                         FITO, category mistake, TLA+, Bell's theorem,
                         indefinite causal order, mutual information
    \item[Related:]      Circumventing FLP (arXiv, 2026),
                         Circumventing CAP (arXiv, 2026),
                         Leibniz Bridge Synthesis (DAE internal)
    \item[License:]      \textcopyright\ 2026 \trauthor, \traffiliation.
                         All rights reserved.
  \end{description}

  \vspace{1.5em}
  \noindent\rule{\linewidth}{0.4pt}

  \vspace{2em}
  \begin{center}
    \normalsize\itshape
    This cover page may be discarded when printing.\\
    The paper begins on the following page.
  \end{center}

  \end{fullwidth}
  \clearpage
  \fi
}

\fancypagestyle{plain}{%
  \fancyhf{}%
  \fancyfoot[L]{\small\textsc{\trshorttitle}}%
  \fancyfoot[C]{\small\thepage\ of \pageref{LastPage}}%
  \ifarxiv\else
  \fancyfoot[R]{\scriptsize\texttt{\trfilename}}%
  \fi
  \ifarxiv\else
  \fancyfoot[L]{\raisebox{-0.8in}{\tiny\texttt{\trversion\ -- \trdate}}}%
  \fancyfoot[R]{\raisebox{-0.8in}{\tiny\texttt{\trfilename}}}%
  \fi
}
\title{Lamport's Arrow of Time:\\The Category Mistake in Logical Clocks}
\author[Paul Borrill]{Paul Borrill, D\AE D\AE LUS}
\date{02026-FEB-25}

\begin{document}
\maketrcover
\setcounter{page}{1}
\maketitle
\placebadges
\thispagestyle{plain}

\daemargintoc

\begin{abstract}
\noindent Lamport's 1978 paper introduced the happens-before relation and logical clocks, freeing distributed systems from dependence on synchronized physical clocks.
This is widely understood as a move away from Newtonian absolute time.
We argue that Lamport's formalism retains a deeper and largely unexamined assumption: that causality induces a globally well-defined directed acyclic graph (DAG) over events---a \emph{forward-in-time-only} (\fito{}) structure that functions as an arrow of time embedded at the semantic level.
Following Ryle's analysis of category mistakes, we show that this assumption conflates an epistemic construct (the logical ordering of messages) with an ontic claim (that physical causality is globally acyclic and monotonic).
We trace this conflation through Shannon's channel model, TLA$^{+}$, Bell's theorem, and the impossibility results of Fischer--Lynch--Paterson and Brewer's CAP theorem.
We then show that special and general relativity permit only local causal structure, and that recent work on indefinite causal order demonstrates that nature admits correlations with no well-defined causal ordering.
We propose that mutual information conservation, rather than temporal precedence, provides a more fundamental primitive for distributed consistency.
\end{abstract}

\FloatBarrier
\section[Introduction]{Introduction}
\label{sec:intro}

In 1978, Leslie Lamport published what would become one of the most cited papers in computer science~\citep{lamport1978}.%
\marginalia[4cm]{Lamport's paper has over 14,000 citations. It is the conceptual ancestor of virtually every consistency protocol in use today, from \paxos{} to Raft to vector clocks.}
Its central contribution was to detach the notion of correctness in a distributed system from synchronized physical clocks.
Instead, Lamport defined a partial order over events---the \emph{happens-before} relation~$\rightarrow$---and showed that a monotonic timestamp function~$C$ could be constructed such that
\begin{equation}
\label{eq:lamport}
a \rightarrow b \quad \Longrightarrow \quad C(a) < C(b).
\end{equation}
The move is widely understood as having freed distributed systems theory from Newtonian absolute time.
We argue that it did not go far enough.

While Lamport eliminated the need for a shared clock, he retained a stronger assumption: that the causal relationships among all events in a distributed system form a globally well-defined directed acyclic graph.%
\marginalia[2cm]{The DAG assumption is so deeply embedded that most distributed systems textbooks do not state it as an assumption at all---it is treated as self-evident.}
This assumption encodes what we call the \emph{Forward-In-Time-Only} (\fito{}) principle---the commitment that causation is irreversible, acyclic, and globally monotonic.
\fito{} functions as an implicit arrow of time, embedded not in the physics of the system but in the semantics of the model.

The purpose of this paper is to make this assumption explicit, to show that it constitutes a \emph{category mistake} in the sense of Ryle~\citep{ryle1949}, and to demonstrate that it is incompatible with the causal structure permitted by both relativity and quantum mechanics.
We further argue that the foundational impossibility results of distributed computing---FLP~\citep{flp1985}, the Two Generals problem, and CAP~\citep{brewer2000,gilbert2002}---are consequences of the \fito{} assumption rather than fundamental physical constraints, and that an information-theoretic primitive based on mutual information conservation offers a path beyond them.

\FloatBarrier
\section[Lamport's Construction]{Lamport's Construction and Its Implicit Commitments}
\label{sec:lamport}

Lamport defines the happens-before relation $\rightarrow$ over events in a distributed system via three rules:
\begin{enumerate}[leftmargin=1.2cm]
    \item If $a$ and $b$ are events in the same process and $a$ comes before $b$ in program order, then $a \rightarrow b$.
    \item If $a$ is the sending of a message and $b$ is the receipt of that message, then $a \rightarrow b$.
    \item Transitivity: if $a \rightarrow b$ and $b \rightarrow c$, then $a \rightarrow c$.
\end{enumerate}
The resulting structure is a strict partial order---equivalently, a directed acyclic graph.
Logical clocks embed this partial order into the natural numbers; vector clocks~\citep{fidge1988,mattern1989} refine the embedding to capture the full partial order rather than merely a consistent extension of it.%
\marginalia{Fidge and Mattern independently invented vector clocks in 1988--89. Unlike Lamport clocks, vector clocks satisfy a biconditional: $a \rightarrow b \iff V(a) < V(b)$. But both still presuppose a DAG.}

These constructions carry three implicit commitments that exceed the mere absence of a synchronized clock:

\begin{definition}[Forward-In-Time-Only (\fito{})]
\label{def:fito}
A distributed systems model satisfies \fito{} if it assumes:
\begin{enumerate}
    \item \textbf{Temporal monotonicity:} every causal chain is mapped to a strictly increasing sequence in some well-ordered set.
    \item \textbf{Asymmetric causation:} if event $a$ can influence event $b$, then $b$ cannot influence $a$; the causal graph contains no directed cycles.
    \item \textbf{Global causal reference:} the partial order $\rightarrow$ is observer-independent; all participants agree on which events are causally related and in which direction.
\end{enumerate}
\end{definition}

\marginalia{Compare with special relativity: two observers in different inertial frames \emph{do not} agree on the temporal ordering of spacelike-separated events. Lamport's third axiom is therefore stronger than what relativity permits.}

Lamport's Rule~(2)---that sending precedes receiving---is the locus of the \fito{} assumption.
It encodes the claim that information flow is unidirectional: the sender's state can influence the receiver's, but not conversely.
As we shall see, this mirrors Shannon's channel model~\citep{shannon1948} and, at a deeper level, the assumption that physical causality admits a single, globally consistent arrow.

\FloatBarrier
\section[The Category Mistake]{The Category Mistake}
\label{sec:category}

Gilbert Ryle introduced the concept of a \emph{category mistake} to describe the error of presenting a fact or entity as if it belonged to a logical category other than its own~\citep{ryle1949}.%
\marginalia{Ryle's famous example: a visitor to Oxford, having been shown the colleges, libraries, and playing fields, asks ``But where is the University?''---confusing an institutional abstraction with a physical location.}

We claim that Lamport's happens-before relation involves an analogous error.
The relation $\rightarrow$ is defined over \emph{message paths}---potential channels of influence between events.
It is an \emph{epistemic} construct: it records what one event \emph{could have known} about another, given the communication topology.
But the embedding into a DAG, and the subsequent use of logical timestamps as a correctness criterion, treats this epistemic ordering as if it were \emph{ontic}---as if it captured the physical causal structure of the system.

Spekkens~\citep{spekkens2007} has demonstrated that many apparently quantum phenomena arise from exactly this type of conflation: treating an epistemic state (knowledge about a system) as if it were an ontic state (the system's physical reality).%
\marginalia{Spekkens' toy model reproduces superposition, entanglement, and no-cloning from a \emph{classical} system with a knowledge-balance principle. The ``mysteries'' disappear once the epistemic--ontic distinction is respected.}
His toy model shows that the distinction between what we \emph{know} and what \emph{is} matters profoundly---a lesson distributed systems has yet to absorb.

The same pattern recurs across the foundations of distributed computing:

\begin{center}
\begin{tabular}{@{}lll@{}}
\toprule
\textbf{Domain} & \textbf{Epistemic (correct)} & \textbf{Ontic (mistake)} \\
\midrule
Quantum mechanics & $\psi$ = knowledge of state & $\psi$ = physical reality \\
Information theory & message = representation & message = physical entity \\
Distributed systems & $\rightarrow$ = potential message path & $\rightarrow$ = physical causation \\
Concurrency & linearization = proof artifact & linearization = physical instant \\
\bottomrule
\end{tabular}
\end{center}

\FloatBarrier
\section[\fito{} in Shannon's Channel]{\fito{} in Shannon's Channel Model}
\label{sec:shannon}

Shannon's mathematical theory of communication~\citep{shannon1948} defines a channel as a unidirectional pipeline:%
\marginalia{Shannon's paper is arguably the founding document of the information age. But its directional channel model---Source $\to$ Encoder $\to$ Channel $\to$ Decoder $\to$ Destination---was a \emph{modeling choice}, not a mathematical necessity.}
\[
\text{Source} \;\longrightarrow\; \text{Encoder} \;\longrightarrow\; \text{Channel} \;\longrightarrow\; \text{Decoder} \;\longrightarrow\; \text{Destination}.
\]
Information flows from source to destination.
The receiver is passive; the additive noise model $Y = X + N$ treats the channel as a forward-only corruption process.

Yet the mathematics of information is symmetric.
Mutual information satisfies
\begin{equation}
\label{eq:mutual}
I(X;Y) = I(Y;X),
\end{equation}
regardless of which variable is ``sent'' and which is ``received.''%
\marginalia{This symmetry is a theorem of probability theory, not an assumption. It holds regardless of any temporal or causal interpretation placed on $X$ and $Y$.}
The asymmetry in Shannon's model is not a consequence of information theory; it is a consequence of the \emph{temporal structure imposed on the model}---the assumption that the source acts before the destination.

Wheeler and Feynman~\citep{wheeler1945} showed that the classical equations of electromagnetism are time-symmetric: both retarded (forward) and advanced (backward) solutions exist.
The apparent arrow---the dominance of retarded radiation---emerges from boundary conditions (the absorber), not from the equations themselves.
Cramer's transactional interpretation~\citep{cramer1986} extends this insight to quantum mechanics.%
\marginalia{Cramer's transactional interpretation models quantum events as a ``handshake'' between a retarded offer wave and an advanced confirmation wave. The transaction is bilateral and atemporal---neither wave is privileged as cause or effect.}

Lamport's Rule~(2) inherits Shannon's \fito{} assumption directly: sending precedes receiving, and no mechanism exists for the receiver's state to constrain the sender's.
The happens-before relation is Shannon's channel, repackaged as a partial order.

\FloatBarrier
\section[\fito{} in TLA$^{+}$]{\fito{} in TLA$^{+}$: The Formal Codification}
\label{sec:tla}

Lamport's 1978 paper introduced the arrow informally.
His \tlaplus{}~\citep{lamport1994tla}---the Temporal Logic of Actions---\emph{formalized} it.%
\marginalia[-0.5cm]{\tlaplus{} is both a specification language and a proof system. It is used at Amazon (AWS), Microsoft, and Intel to verify distributed protocols. Its adoption makes the \fito{} assumption operationally consequential, not merely theoretical.}
\tlaplus{} is built on linear temporal logic~\citep{pnueli1977} and encodes \fito{} at every level of its semantics.

\subsection{Unprimed/Primed: \fito{} in the Syntax}

In \tlaplus{}, a state is an assignment of values to variables.
An \emph{action} is a predicate over pairs of states: the \emph{current} state (unprimed variables $x$) and the \emph{next} state (primed variables $x'$).
The syntax itself enforces unidirectionality---there is no mechanism to refer to a \emph{previous} state.%
\marginalia[-1cm]{Compare with physics: the Euler--Lagrange equations of classical mechanics are second-order and time-symmetric. They constrain trajectories, not directions. \tlaplus{}'s unprimed/primed architecture imposes a direction that the physics does not require.}
A behavior is an infinite sequence of states $s_0, s_1, s_2, \ldots$---always forward, never backward.

\subsection{Temporal Operators: Forward-Only Windows}

\tlaplus{}'s temporal operators are all forward-looking:
\begin{itemize}
    \item $\Box P$ (``always $P$''): $P$ holds in every future state.
    \item $\Diamond P$ (``eventually $P$''): $P$ holds in some future state.
    \item $P \leadsto Q$ (``$P$ leads to $Q$''): whenever $P$ holds, $Q$ eventually holds \emph{afterward}.
\end{itemize}
There is no past-tense operator.%
\marginalia[-1cm]{Some temporal logics (e.g., PLTL) include past operators: ``$P$ was true at some point in the past.'' Lamport deliberately excluded them from TLA. The exclusion is a design choice, not a logical necessity.}
The logic cannot express ``$Q$ was caused by some earlier $P$'' except by forward implication from $P$.
This is \fito{} expressed as proof methodology.

\subsection{Stuttering Equivalence: Direction Without Reversal}

Two behaviors that differ only by finite repetitions of states (``stuttering steps'') satisfy the same \tlaplus{} formulas.
Stuttering equivalence is a powerful refinement tool---it allows high-level and low-level specifications to be compared without matching step-for-step.%
\marginalia[-1cm]{Stuttering is the key to \tlaplus{}'s compositional proof method. But notice: a stutter \emph{repeats} the current state. It does not revisit a previous state. The operation is idempotent, not reversible.}

But stuttering is strictly forward: it inserts copies of the current state, never copies of a prior state.
The equivalence preserves temporal direction.
Reversal---visiting $s_3$ after $s_5$---is not stuttering; it is undefined.

\subsection{Safety and Liveness: A Temporal Asymmetry}

Every \tlaplus{} property decomposes into safety (``nothing bad happens'') and liveness (``something good eventually happens'')~\citep{abadi1988}.%
\marginalia[-1.5cm]{Alpern and Schneider (1985) proved that every linear temporal property is the intersection of a safety and a liveness property. This decomposition is \emph{itself} temporally asymmetric: safety constrains prefixes; liveness constrains suffixes.}
Safety properties are \emph{closed} sets---violated by finite prefixes.
Liveness properties are \emph{dense} sets---requiring infinite futures.

There is no dual notion of ``backward liveness'' (something good happened in the past) or ``backward safety'' (nothing bad happened before).
The decomposition presupposes a distinguished temporal direction.

\subsection{Fairness: Constraining the Future, Not the Past}

\tlaplus{}'s fairness conditions---weak fairness ($\text{WF}$) and strong fairness ($\text{SF}$)---constrain what \emph{must eventually happen} given what is \emph{enabled now}:%
\marginalia[-1.5cm]{Without fairness, a \tlaplus{} specification permits behaviors where an enabled action is ignored forever. Fairness rules this out---but it can only look forward: ``if continuously enabled, then eventually taken.'' It cannot look backward.}
\begin{itemize}
    \item $\text{WF}_f(A)$: if action $A$ is continuously enabled, it must eventually be taken.
    \item $\text{SF}_f(A)$: if action $A$ is infinitely often enabled, it must eventually be taken.
\end{itemize}
Fairness conditions use future information to constrain present behavior.
They cannot use past information to constrain the present---this is machine closure, and it enforces \fito{} at the level of proof obligations.

\subsection{The 1978--1994 Trajectory}

Lamport's intellectual trajectory from the 1978 paper to \tlaplus{} in 1994 is revealing.%
\marginalia[-1.5cm]{This is not a criticism of Lamport's engineering judgment. \tlaplus{} is a superb tool for reasoning about forward-time systems. The question is whether \emph{all} systems are forward-time systems---and physics says they are not.}
The happens-before relation was an informal insight; \tlaplus{} is its formalization.
The \fito{} assumption, implicit in 1978, became the \emph{axiom} of 1994.
What was a modeling choice became a logical foundation.

\FloatBarrier
\section[\fito{} and Impossibility Theorems]{\fito{} and the Impossibility Theorems}
\label{sec:impossibility}

The three canonical impossibility results of distributed computing---the Two Generals problem, FLP~\citep{flp1985}, and CAP~\citep{brewer2000,gilbert2002}---are typically presented as fundamental physical constraints on what distributed systems can achieve.
We argue instead that they are \emph{consequences of the \fito{} assumption}.%
\marginalia[-1cm]{This is a strong claim. We are not saying the theorems are wrong. We are saying they are theorems about a \emph{particular model}---one that assumes forward-only information flow---and that the model is not the only possible one.}

The Two Generals problem demonstrates that no finite number of unreliable unidirectional messages can guarantee common knowledge.
But the assumption of unidirectionality is precisely \fito{}: each message travels forward in time, and the only way to confirm receipt is to send another forward-traveling message, leading to an infinite regress.

FLP proves that no deterministic asynchronous protocol can guarantee consensus in the presence of even a single crash failure.
The proof exploits the indistinguishability, from any single process's perspective, of a crashed process and a slow one.%
\marginalia{The FLP proof constructs a bivalent initial configuration and shows that every step preserves bivalence. The construction depends on the fact that processes cannot ``reach back'' to determine a peer's state---a direct consequence of \fito{}.}
This indistinguishability arises because information propagates only forward---a process cannot ``reach back'' to determine whether a peer has failed or is merely delayed.

CAP shows that a distributed system cannot simultaneously guarantee consistency, availability, and partition tolerance.
Consistency, in this context, requires maintaining a globally agreed causal order across partitioned subsystems---which is precisely the enforcement of a global DAG under conditions where the communication topology no longer supports it.

In each case, the impossibility follows from the combination of asynchrony and \fito{}.
If information flow were not constrained to be unidirectional---if, for instance, the causal structure of the system were treated as a bilateral, transactional exchange rather than a sequence of one-way messages---the premises of the impossibility proofs would not hold in their present form.

\FloatBarrier
\section[Relativity]{Relativity and Local Causal Structure}
\label{sec:relativity}

Special relativity eliminates absolute simultaneity: for spacelike-separated events, temporal order is frame-dependent.%
\marginalia{Lamport explicitly cited the relativistic motivation for his construction. But he extracted only the partial ordering (no absolute simultaneity) while discarding the deeper lesson: causal structure is \emph{local}, not global.}
This much is well known, and Lamport explicitly cites the relativistic motivation for his construction~\citep{lamport1978}.

However, relativity does more than eliminate global simultaneity.
In Minkowski spacetime, causal structure is \emph{local}: only events within each other's light cones stand in a definite causal relation.
Spacelike-separated events are not merely ``unordered''---they are \emph{causally disconnected}.
No global DAG exists over the full event space; the causal structure is a local partial order defined by the metric.

Lamport's model is compatible with relativity in a narrow sense: it does not require a global clock.
But it presumes something stronger---that all message-based interactions can be embedded in a single, globally consistent partial order.%
\marginalia{In relativistic terms, Lamport assumes that all communication occurs within a single connected causal diamond. For a datacenter this is reasonable. For a planetary-scale or cislunar network, it is not.}

General relativity further complicates the picture.
In curved spacetime, the causal structure is dynamical---determined by the distribution of matter and energy.
Closed timelike curves are permitted solutions to the Einstein field equations.
The causal structure of the universe is not a fixed DAG imposed from above; it is a dynamical, geometry-dependent, and potentially topology-changing feature of the spacetime manifold.

Lamport eliminates the Newtonian clock.
But he retains the Newtonian assumption that causal order is global, acyclic, and observer-independent.
Relativity tells us that none of these properties is guaranteed.

\FloatBarrier
\section[Indefinite Causal Order]{Indefinite Causal Order}
\label{sec:ico}

Recent work in quantum foundations demonstrates that causal order itself need not be definite.
Oreshkov, Costa, and Brukner~\citep{oreshkov2012} introduced the \emph{process matrix} formalism, which describes quantum correlations between operations without presupposing a background causal structure.%
\marginalia{A process matrix $W$ acts on the tensor product of input and output spaces of local laboratories. If $W$ can be decomposed as a probabilistic mixture of definite causal orders, it is \emph{causally separable}; otherwise, the operations exist in a genuine superposition of causal orders.}

Chiribella, D'Ariano, Perinotti, and Valiron~\citep{chiribella2013} independently constructed the \emph{quantum switch}, in which a control qubit determines whether operations $A$ and $B$ are applied in the order $A \circ B$ or $B \circ A$.
When the control is in superposition, neither order is realized:
\begin{equation}
\label{eq:switch}
\frac{1}{\sqrt{2}}\bigl(\lvert 0 \rangle \otimes A \circ B + \lvert 1 \rangle \otimes B \circ A\bigr).
\end{equation}
No assignment of timestamps to $A$ and $B$ is consistent with the output.
Happens-before is not merely unknown; it is \emph{undefined}.%
\marginalia{This is the key point for distributed systems: the quantum switch is not a case of missing information about causal order. It is a case where \emph{no causal order exists}. Lamport's formalism has no representation for this.}

Hardy~\citep{hardy2005,hardy2007} anticipated this development by proposing a framework for probabilistic theories with \emph{dynamic} causal structure, motivated by quantum gravity.
His \emph{causaloid} formalism tracks causal dependencies across regions of a system without assuming a fixed background ordering.

\subsection{Experimental Verification}

Indefinite causal order is not merely a theoretical possibility.
Rubino et al.~\citep{rubino2017} provided the first experimental verification using a photonic quantum switch, measuring causal witnesses that violate bounds achievable by any causally definite process.
Goswami et al.~\citep{goswami2018} demonstrated indefinite causal order at the single-photon level.
Procopio et al.~\citep{procopio2015} demonstrated experimental superposition of gate orders.%
\marginalia{These are laboratory experiments, not thought experiments. The causal witnesses measured by Rubino et al.\ violate bounds that \emph{any} causally definite process must satisfy---analogous to Bell inequality violations ruling out local hidden variables.}

The implications for Lamport's framework are direct: nature permits physical processes for which no DAG of events exists.
Logical clocks, vector clocks, and any formalism that assumes a well-defined partial order of events are \emph{inapplicable in principle} to systems exhibiting indefinite causal order.

\FloatBarrier
\section[Bell's Theorem]{Bell's Theorem and the \fito{} Conditional}
\label{sec:bell}

Bell's 1964 theorem~\citep{bell1964} is often summarized as proving that quantum mechanics is nonlocal.
More precisely, it proves that \emph{if} one assumes (i)~locality, (ii)~realism, and (iii)~a forward-in-time-only causal structure, \emph{then} the resulting correlations satisfy inequalities that quantum mechanics violates.%
\marginalia{The logical structure of Bell's theorem is: $(L \land R \land \text{\fito{}}) \Rightarrow \text{Bell inequality}$. Nature violates the inequality. The standard response abandons $L$ (locality). But one could equally abandon \fito{}.}

The logical structure is a conditional.
\fito{} is one of the premises, not a conclusion.
The standard response---abandon locality---is not the only option.
One may instead abandon \fito{}: retrocausal models~\citep{cramer1986} and the transactional interpretation restore locality by permitting causal influence from future to past.
Wood and Spekkens~\citep{wood2015} have shown that any \fito{}-respecting causal model of Bell violations requires \emph{fine-tuning}---a systematic and unexplained coincidence in the causal parameters.%
\marginalia{Fine-tuning in causal models is analogous to fine-tuning in physics: it suggests that the model is missing something. Dropping \fito{} dissolves the fine-tuning problem entirely.}
Dropping \fito{} dissolves the fine-tuning problem.

For distributed systems, the lesson is this: Bell's theorem does not validate \fito{}.
It shows that \fito{}, combined with locality and realism, is \emph{insufficient to account for observed physics}.
A distributed systems theory built on \fito{} inherits this insufficiency.

\FloatBarrier
\section[Category-Theoretic Frameworks]{Category-Theoretic Frameworks Beyond the DAG}
\label{sec:category_theory}

If the DAG is not fundamental, what replaces it?
Category-theoretic approaches to quantum mechanics and causality offer one answer.

Coecke and Kissinger~\citep{coecke2017,kissinger2017} develop \emph{categorical quantum mechanics} (CQM), in which processes (morphisms) and their composition are primitive.%
\marginalia{In CQM, a process is a morphism in a monoidal category. Composition is sequential; tensor product is parallel. Causal structure emerges from the algebra, not from a pre-imposed graph. The DAG is one possible algebra---but not the only one.}
Causal structure emerges from the composition rules and symmetries of a monoidal category, rather than being imposed as a global DAG.
CQM naturally accommodates indefinite causal order: the composition of morphisms does not require a pre-defined total or partial order.

Abramsky~\citep{abramsky2023} extends this approach with a \emph{topological} treatment of causality, in which causal structure is not a combinatorial object (a graph) but a topological one---admitting continuous deformations between regions of definite and indefinite order.
His earlier work with Brandenburger~\citep{abramsky2011} established a sheaf-theoretic framework for contextuality that applies directly to causal structure: the local consistency of causal orders need not extend to global consistency, just as local measurement contexts in quantum mechanics need not admit a global hidden-variable model.%
\marginalia{Sheaf theory formalizes exactly the pattern we see in distributed systems: local views are consistent, but they cannot always be patched together into a globally consistent picture. Abramsky and Brandenburger proved that this is a \emph{topological} obstruction, not merely a practical difficulty.}

Lal~\citep{lal2012} formalizes causal structure within CQM as a categorical construct that subsumes definite, dynamical, and indefinite orderings as special cases.

These frameworks share a crucial feature: they do not presuppose a global DAG.
Lamport's happens-before is a special case---the case in which causal structure is both definite and globally consistent.
The categorical approach reveals this as one point in a much larger space of possibilities.

\FloatBarrier
\section[Information Symmetry]{Information Symmetry as Primitive}
\label{sec:information}

If temporal precedence is not fundamental, what is?
We propose that the primitive for distributed consistency should be \emph{symmetric mutual information alignment} rather than happens-before ordering.%
\marginalia{This is the constructive proposal of the paper. The preceding sections show that \fito{} is a category mistake. This section sketches what might replace it.}

Mutual information $I(X;Y) = I(Y;X)$ is inherently symmetric.
It measures the reduction in uncertainty about one variable given knowledge of the other, without privileging either variable as ``cause'' or ``effect.''
Landauer's principle~\citep{landauer1961} and Bennett's analysis of reversible computation~\citep{bennett1982} establish that information and thermodynamics are linked---but the link is through entropy production, not through a pre-given causal arrow.%
\marginalia{Landauer (1961): erasing one bit costs $kT \ln 2$ of energy. Bennett (1982): computation can be thermodynamically reversible if no information is erased. The arrow of time in computation is the arrow of \emph{erasure}, not the arrow of \emph{causation}.}
The thermodynamic arrow of time is itself emergent, a consequence of boundary conditions and the second law, not a fundamental feature of the microscopic dynamics.

This suggests a reframing:
\begin{equation}
\text{Correctness} \;\neq\; \text{Temporal Precedence}.
\end{equation}
Instead:
\begin{equation}
\text{Correctness} \;=\; \text{Symmetric Mutual Information Conservation}.
\end{equation}

In this view, a distributed transaction is correct not because events occur in the right order, but because the mutual information between interacting subsystems is conserved across the interaction boundary.
Failure is not a violation of temporal ordering; it is an \emph{epistemic mismatch}---a divergence in what the subsystems know about each other.

Rovelli's relational quantum mechanics~\citep{rovelli1996} provides a physical model for this perspective: properties exist only relative to other systems, and ``state'' is a summary of one system's information about another.%
\marginalia{Rovelli: ``The world is not a collection of things, it is a collection of events\ldots'' If state is relational, then consistency is a property of \emph{relationships}, not of \emph{orderings}.}
The thermal time hypothesis of Connes and Rovelli~\citep{connes1994} further suggests that time itself is an emergent parameter of the informational state, not a background against which information evolves.

\FloatBarrier
\section[Logical Clocks as Coordinates]{Logical Clocks as Coordinate Systems}
\label{sec:coordinates}

In general relativity, coordinate systems are representational tools, not features of physical reality.
The physics is contained in the metric tensor and the curvature, not in the choice of coordinates.%
\marginalia{Einstein spent years struggling with coordinate systems before arriving at general covariance: the principle that the laws of physics take the same form in all coordinate systems. The content of general relativity is \emph{coordinate-free}.}

We propose that logical clocks occupy an analogous role in distributed systems.
They are \emph{coordinate systems} imposed on an underlying causal structure---useful for computation and proof, but not ontologically fundamental.

Lamport clocks embed the partial order into $\mathbb{N}$.
Vector clocks~\citep{fidge1988,mattern1989} embed it into $\mathbb{N}^n$.
Both embeddings presuppose that the underlying structure is a DAG---that there \emph{exists} a partial order to embed.
If the causal structure is indefinite, or if it is relational and observer-dependent, then the embedding is not merely approximate; it is a \emph{misrepresentation of the ontology}.

This is the category mistake made precise: logical clocks treat an epistemic coordinate system as if it were an ontic feature of the distributed system.

\FloatBarrier
\section[Discussion]{Discussion}
\label{sec:discussion}

We are not arguing that Lamport's work is without value.
The happens-before relation and logical clocks were a genuine advance: they showed that distributed correctness does not require synchronized physical clocks, and they provided practical tools that remain in wide use nearly fifty years later.%
\marginalia{Lamport received the Turing Award in 2013, in large part for this work. The award was deserved. Our argument is that the advance was incomplete, not that it was wrong.}

Our argument is that the advance was incomplete.
Lamport removed the clock but retained the arrow.
He replaced Newtonian absolute time with a different absolute: a globally well-defined, acyclic, observer-independent causal order.
This is weaker than Newton, but it is still stronger than what physics requires.

Relativity requires only local causal structure.
Quantum mechanics permits indefinite causal order.
The thermodynamic arrow of time is emergent, not fundamental.
Bell's theorem, read correctly, shows that \fito{} is a \emph{problematic assumption}, not a validated one.

The impossibility theorems of distributed computing---FLP, Two Generals, CAP---are theorems about \fito{} systems.
They tell us what cannot be done \emph{if} information flows only forward.
They do not tell us what cannot be done in general.
Recognizing this distinction opens a design space that the current theoretical framework forecloses.

A distributed systems theory built on information symmetry rather than temporal precedence---on mutual information conservation rather than happens-before ordering---would not be subject to the same impossibility results, because it would not make the same assumptions.
Whether such a theory can be made practical is an engineering question.
That it is \emph{conceptually coherent} is what this paper has aimed to establish.

\FloatBarrier
\section[Conclusion]{Conclusion}
\label{sec:conclusion}

Lamport's 1978 paper is a landmark.
But its happens-before relation embeds a hidden arrow of time---the Forward-In-Time-Only assumption---that is stronger than anything physics requires and incompatible with the indefinite causal order that quantum mechanics permits.

This arrow is a category mistake: it treats an epistemic ordering of message paths as if it were the ontic causal structure of physical reality.
Relativity, quantum foundations, and information theory all point toward a deeper primitive---one in which time and causal order are emergent rather than fundamental, and in which mutual information, not temporal precedence, serves as the basis for consistency.

Logical clocks are coordinate systems, not physics.
Lamport's arrow of time is not the arrow of time.%
\marginalia{``Lamport removed the clock but retained the arrow.'' This sentence summarizes the entire paper.}

\section*{Acknowledgments}

The author acknowledges the use of AI-assisted tools for literature review and structural refinement.
All interpretations, claims, and conclusions are solely the responsibility of the author.


\bibliographystyle{unsrtnat}
\bibliography{references}

@article{lamport1978,
  author    = {Leslie Lamport},
  title     = {Time, Clocks, and the Ordering of Events in a Distributed System},
  journal   = {Communications of the ACM},
  volume    = {21},
  number    = {7},
  pages     = {558--565},
  year      = {1978},
}

@inproceedings{fidge1988,
  author    = {Colin J. Fidge},
  title     = {Timestamps in Message-Passing Systems That Preserve the Partial Ordering},
  booktitle = {Proceedings of the 11th Australian Computer Science Conference},
  pages     = {56--66},
  year      = {1988},
}

@inproceedings{mattern1989,
  author    = {Friedemann Mattern},
  title     = {Virtual Time and Global States of Distributed Systems},
  booktitle = {Proceedings of the International Workshop on Parallel and Distributed Algorithms},
  pages     = {215--226},
  year      = {1989},
}

@article{flp1985,
  author    = {Michael J. Fischer and Nancy A. Lynch and Michael S. Paterson},
  title     = {Impossibility of Distributed Consensus with One Faulty Process},
  journal   = {Journal of the ACM},
  volume    = {32},
  number    = {2},
  pages     = {374--382},
  year      = {1985},
}

@inproceedings{brewer2000,
  author    = {Eric A. Brewer},
  title     = {Towards Robust Distributed Systems},
  booktitle = {Proceedings of the 19th ACM Symposium on Principles of Distributed Computing},
  year      = {2000},
  note      = {Keynote address},
}

@article{gilbert2002,
  author    = {Seth Gilbert and Nancy Lynch},
  title     = {Brewer's Conjecture and the Feasibility of Consistent, Available, Partition-Tolerant Web Services},
  journal   = {ACM SIGACT News},
  volume    = {33},
  number    = {2},
  pages     = {51--59},
  year      = {2002},
}

@article{shannon1948,
  author    = {Claude E. Shannon},
  title     = {A Mathematical Theory of Communication},
  journal   = {The Bell System Technical Journal},
  volume    = {27},
  number    = {3},
  pages     = {379--423},
  year      = {1948},
}

@article{rovelli1996,
  author    = {Carlo Rovelli},
  title     = {Relational Quantum Mechanics},
  journal   = {International Journal of Theoretical Physics},
  volume    = {35},
  number    = {8},
  pages     = {1637--1678},
  year      = {1996},
  eprint    = {quant-ph/9609002},
  archiveprefix = {arXiv},
}

@article{connes1994,
  author    = {Alain Connes and Carlo Rovelli},
  title     = {Von {N}eumann Algebra Automorphisms and Time-Thermodynamics Relation in Generally Covariant Quantum Theories},
  journal   = {Classical and Quantum Gravity},
  volume    = {11},
  number    = {12},
  pages     = {2899--2917},
  year      = {1994},
}

@article{wheeler1945,
  author    = {John Archibald Wheeler and Richard P. Feynman},
  title     = {Interaction with the Absorber as the Mechanism of Radiation},
  journal   = {Reviews of Modern Physics},
  volume    = {17},
  number    = {2--3},
  pages     = {157--181},
  year      = {1945},
}

@article{cramer1986,
  author    = {John G. Cramer},
  title     = {The Transactional Interpretation of Quantum Mechanics},
  journal   = {Reviews of Modern Physics},
  volume    = {58},
  number    = {3},
  pages     = {647--688},
  year      = {1986},
}

@article{oreshkov2012,
  author    = {Ognyan Oreshkov and Fabio Costa and \v{C}aslav Brukner},
  title     = {Quantum Correlations with No Causal Order},
  journal   = {Nature Communications},
  volume    = {3},
  pages     = {1092},
  year      = {2012},
}

@article{chiribella2013,
  author    = {Giulio Chiribella and Giacomo Mauro D'Ariano and Paolo Perinotti and Benoit Valiron},
  title     = {Quantum Computations without Definite Causal Structure},
  journal   = {Physical Review A},
  volume    = {88},
  number    = {2},
  pages     = {022318},
  year      = {2013},
  eprint    = {0912.0195},
  archiveprefix = {arXiv},
}

@article{hardy2005,
  author    = {Lucien Hardy},
  title     = {Probability Theories with Dynamic Causal Structure: A New Framework for Quantum Gravity},
  year      = {2005},
  eprint    = {gr-qc/0509120},
  archiveprefix = {arXiv},
}

@article{hardy2007,
  author    = {Lucien Hardy},
  title     = {Towards Quantum Gravity: A Framework for Probabilistic Theories with Non-Fixed Causal Structure},
  journal   = {Journal of Physics A},
  volume    = {40},
  pages     = {3081--3099},
  year      = {2007},
  eprint    = {gr-qc/0608043},
  archiveprefix = {arXiv},
}

@article{rubino2017,
  author    = {Giulia Rubino and Lee A. Rozema and Adrien Feix and Mateus Ara\'{u}jo and Jonas M. Zeuner and Lorenzo M. Procopio and \v{C}aslav Brukner and Philip Walther},
  title     = {Experimental Verification of an Indefinite Causal Order},
  journal   = {Science Advances},
  volume    = {3},
  number    = {3},
  pages     = {e1602589},
  year      = {2017},
}

@article{goswami2018,
  author    = {K. Goswami and C. Giarmatzi and M. Kewming and F. Costa and C. Branciard and J. Romero and A. G. White},
  title     = {Indefinite Causal Order in a Quantum Switch},
  journal   = {Physical Review Letters},
  volume    = {121},
  pages     = {090503},
  year      = {2018},
}

@article{procopio2015,
  author    = {Lorenzo M. Procopio and Amir Moqanaki and Marco Ara\'{u}jo and Fabio Costa and Irati Alonso Calafell and Emma G. Dowd and Deny R. Hamel and Lee A. Rozema and \v{C}aslav Brukner and Philip Walther},
  title     = {Experimental Superposition of Orders of Quantum Gates},
  journal   = {Nature Communications},
  volume    = {6},
  pages     = {7913},
  year      = {2015},
}

@book{coecke2017,
  author    = {Bob Coecke and Aleks Kissinger},
  title     = {Picturing Quantum Processes: A First Course in Quantum Theory and Diagrammatic Reasoning},
  publisher = {Cambridge University Press},
  year      = {2017},
}

@article{kissinger2017,
  author    = {Bob Coecke and Aleks Kissinger},
  title     = {Categorical Quantum Mechanics {I}: Causal Quantum Processes},
  year      = {2017},
  eprint    = {1510.05468},
  archiveprefix = {arXiv},
  note      = {In \emph{Categories for the Working Philosopher}, ed.\ E.~Landry, Oxford University Press},
}

@article{abramsky2023,
  author    = {Samson Abramsky},
  title     = {The Topology of Causality},
  year      = {2023},
  eprint    = {2303.07148},
  archiveprefix = {arXiv},
}

@article{abramsky2011,
  author    = {Samson Abramsky and Adam Brandenburger},
  title     = {The Sheaf-Theoretic Structure of Non-Locality and Contextuality},
  journal   = {New Journal of Physics},
  volume    = {13},
  pages     = {113036},
  year      = {2011},
}

@phdthesis{lal2012,
  author    = {Raymond Lal},
  title     = {Causal Structure in Categorical Quantum Mechanics},
  school    = {University of Oxford},
  year      = {2012},
}

@book{ryle1949,
  author    = {Gilbert Ryle},
  title     = {The Concept of Mind},
  publisher = {Hutchinson},
  year      = {1949},
}

@article{spekkens2007,
  author    = {Robert W. Spekkens},
  title     = {Evidence for the Epistemic View of Quantum States: A Toy Theory},
  journal   = {Physical Review A},
  volume    = {75},
  pages     = {032110},
  year      = {2007},
}

@article{wood2015,
  author    = {Christopher J. Wood and Robert W. Spekkens},
  title     = {The Lesson of Causal Discovery Algorithms for Quantum Correlations: Causal Explanations of {B}ell-Inequality Violations Require Fine-Tuning},
  journal   = {New Journal of Physics},
  volume    = {17},
  pages     = {033002},
  year      = {2015},
}

@article{bell1964,
  author    = {John S. Bell},
  title     = {On the {E}instein-{P}odolsky-{R}osen Paradox},
  journal   = {Physics},
  volume    = {1},
  number    = {3},
  pages     = {195--200},
  year      = {1964},
}

@article{lamport1994tla,
  author    = {Leslie Lamport},
  title     = {The Temporal Logic of Actions},
  journal   = {ACM Transactions on Programming Languages and Systems},
  volume    = {16},
  number    = {3},
  pages     = {872--923},
  year      = {1994},
}

@inproceedings{abadi1988,
  author    = {Mart\'{\i}n Abadi and Leslie Lamport},
  title     = {The Existence of Refinement Mappings},
  booktitle = {Proceedings of the Third Annual IEEE Symposium on Logic in Computer Science (LICS)},
  year      = {1988},
}

@inproceedings{pnueli1977,
  author    = {Amir Pnueli},
  title     = {The Temporal Logic of Programs},
  booktitle = {Proceedings of the 18th IEEE Symposium on Foundations of Computer Science},
  pages     = {46--57},
  year      = {1977},
}

@article{landauer1961,
  author    = {Rolf Landauer},
  title     = {Irreversibility and Heat Generation in the Computing Process},
  journal   = {IBM Journal of Research and Development},
  volume    = {5},
  number    = {3},
  pages     = {183--191},
  year      = {1961},
}

@article{bennett1982,
  author    = {Charles H. Bennett},
  title     = {The Thermodynamics of Computation---A Review},
  journal   = {International Journal of Theoretical Physics},
  volume    = {21},
  number    = {12},
  pages     = {905--940},
  year      = {1982},
}

\end{document}